\documentclass[11pt,a4paper]{article} 
\usepackage{jheppub}

\usepackage{epsfig}
\usepackage{latexsym}
\usepackage{graphicx}
\usepackage{epstopdf}  
\usepackage[justification=RaggedRight]{caption}
\usepackage{amssymb}
\usepackage{subfig}
\usepackage{color}
\usepackage{axodraw}
\usepackage{multirow}
\usepackage{color}
\usepackage{rotating}

\usepackage{hyperref} 

\usepackage{hyperref}
\hypersetup{
    colorlinks,%
    citecolor=blue,%
    filecolor=blue,%
    linkcolor=blue,%
    urlcolor=blue
}

\def\be{\begin{equation}}
\def\ee{\end{equation}}

\newcommand{\vev}[1]{\left\langle #1\right\rangle}

\newcommand{\bwt}{\begin{widetext}}
\newcommand{\ewt}{\end{widetext}}
\newcommand{\bdm}{\begin{displaymath}}
\newcommand{\edm}{\end{displaymath}}
\newcommand{\bea}{\begin{eqnarray}}
\newcommand{\eea}{\end{eqnarray}}
\newcommand{\nn}{\nonumber}

\def\eq#1{{Eq.~(\ref{#1})}}
\def\eqs#1#2{{Eqs.~(\ref{#1})--(\ref{#2})}}
\def\fig#1{{Fig.~\ref{#1}}}

\def\Table#1{{Table~\ref{#1}}}

\def\sect#1{{Sect.~\ref{#1}}}

\def\app#1{{Appendix~\ref{#1}}}

\def\vev#1{\left\langle #1\right\rangle}

\def\det{\mbox{det}\,}

\begin{document}

\title{Gravitino Dark Matter in Tree Level Gauge Mediation with and without R-parity}

\author[a]{Giorgio Arcadi,} 
\author[a]{Luca Di Luzio,%
} 
\author[b]{Marco Nardecchia} 

\affiliation[a]{SISSA/ISAS and INFN, Sezione di Trieste, \\ 
Via Bonomea 265, I-34136 Trieste, Italy.} 
\affiliation[b]{$CP^3$-Origins and Danish Institute for Advanced Study DIAS, 
University of Southern Denmark, \\ 
Campusvej 55, DK-5230, Odense M, Denmark.} 

\emailAdd{arcadi@sissa.it} 
\emailAdd{diluzio@sissa.it} 
\emailAdd{nardecchia@cp3-origins.net}

\abstract{
We investigate the cosmological aspects of Tree Level Gauge Mediation, a recently proposed mechanism in which 
the breaking of supersymmetry is communicated to the soft scalar masses by extra gauge interactions at the tree level. 
Embedding the mechanism in a Grand Unified Theory and requiring the observability of sfermion 
masses at the Large Hadron Collider, it follows that the Lightest Supersymmetric Particle is a gravitino with a mass of the order of $10$ GeV. 
The analysis in the presence of R-parity shows that a typical Tree Level Gauge Mediation spectrum 
leads to an overabundance of the Dark Matter relic density and a tension with the 
constraints from Big Bang Nucleosynthesis. This suggests to relax the exact conservation of the R-parity. 
The underlying $SO(10)$ Grand Unified Theory together with the bounds from proton decay 
provide a rationale for considering only bilinear R-parity violating operators. 
We finally analyze the cosmological implications of this setup by identifying the 
phenomenologically viable regions of the parameter space.
} 
\keywords{Cosmology of Theories beyond the SM, Supersymmetry Phenomenology, GUT.}

\maketitle

\section{Introduction}

Low-energy supersymmetry (SUSY) provides an appealing framework for the origin of the Dark Matter (DM) component of the energy density of the Universe. 
In the Minimal Supersymmetric Standard Model (MSSM) only two particle states stand out for their phenomenological viability: the neutralino and the gravitino.

In this work we investigate the case where the DM candidate is the gravitino with a mass in a specific range dictated by Tree Level Gauge Mediation (TGM)~\cite{Nardecchia:2009ew,Nardecchia:2009nh,Monaco:2011fe}, a recently proposed mechanism in which the breaking of SUSY is communicated to the soft scalar masses by extra gauge interactions at the tree level. 

The mass of the gravitino is particularly sensitive to the mechanism of SUSY breaking and its mediation 
to the visible sector. 
In a large class of models SUSY is broken by the F-term of a chiral superfield $\vev{Z} = F \theta^2$ and the breaking is 
communicated to the Standard Model (SM) chiral superfields at the mediation scale $M$. 
The two most popular mechanisms of mediation of SUSY breaking are gravity and (loop) gauge mediation.

In the former case the scale $M$ coincides with the Planck scale, $M_P$. 
Then the soft terms and the gravitino mass are expected to be of the order of $F / M_{P}$,
implying that the gravitino is not always the LSP and its mass is comparable with $m_{\rm soft} \sim 1 \ \rm{TeV}$.
On the other hand, in theories based on loop gauge mediation the soft scale is
\be
m_{\rm soft} \sim \frac{\alpha}{4\pi}\frac{F}{M} \, ,
\ee
depending on the two parameters $F$ and $M$. In this case the gravitino is always the LSP 
with a mass ranging from $\mathcal{O}(\rm{eV})$ to $\mathcal{O}(\rm{GeV})$. 

In this paper we investigate the cosmology of the alternative scenario given by TGM.
In such a case the chiral superfield $Z$ is a SM singlet with charge $X_Z$ under $U(1)_X$, 
which is an extra abelian gauge symmetry broken at the scale $M_X$.
The MSSM superfields $Q$ carry a non trivial charge $X_Q$ under $U(1)_X$. 
Integrating out the heavy vector superfield $V_X$ related to the $U(1)_X$ factor,
\begin{figure}[!h]
\begin{center}
\begin{picture}(240,70)(30,30)
\SetWidth{1.5}
\Photon(120,50)(180,50){4}{6}
\SetWidth{1.5}
\Line(90,75)(120,50)
\Line(90,25)(120,50)
\Line(210,75)(180,50)
\Line(210,25)(180,50)
\SetWidth{0.5}
\Text(86,80)[]{ $Z^{\dagger}$ }
\Text(83, 28)[]{ $Z$ }
\Text(150,62)[]{ $V_X$ }
\Text(218,80)[]{ $Q^{\dagger}$ }
\Text(220, 28)[]{ $Q $ }
\end{picture}
\end{center}
\caption{TGM super-graph generating the bilinear soft masses.}
\label{fig:diagram}
\end{figure}
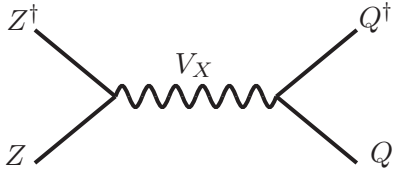
soft masses are induced through the F-term $\vev{Z} = F \theta^2$ and read
\begin{equation}
\label{soft1}
(\tilde{m}^2_{Q})_\text{tree}= g^2_X X_Q X_Z \frac{F^2}{M_X^2} \, ,
\end{equation}
where $g_X$ is the gauge coupling relative to $U(1)_X$.

The gravitino mass is then related to $m_{\rm soft}$ by the simple relation
\be
\label{eq:general32}
m_{3/2} \sim m_{\rm soft} \frac{M_X}{M_{P}} \, .
\ee
It is natural to embed the the enlarged  $G_{SM} \otimes U(1)_X $ group 
into a rank-5 Grand Unified Theory (GUT) such as $SO(10)$. 
Hence we assume that the SUSY breaking is communicated at the $SO(10) \to G_{SM}$ breaking scale $M_G \sim 10^{16} \ \rm{GeV}$. 
We note a few interesting features of this model:
\begin{itemize}
\item the sfermions masses are flavour universal as in (loop) gauge mediation, thus solving the supersymmetric flavour problem.
\item from~\eq{soft1} we can see that the ratios among the soft masses $(\tilde{m}^2_{Q})_\text{tree}$ depend just on the choice of the $U(1)_X$ charges, 
making the model predictive and testable at the LHC.
\item given $M_X=M_G$,~\eq{eq:general32} implies that the gravitino mass is of the order of 10 
GeV\footnote{Such value for the gravitino mass is not typically accessible in standard (loop) gauge mediation. 
On the other hand LSP gravitinos of $\mathcal{O}(10 \ \rm{GeV})$ are also achievable in scenarios
like gaugino mediated SUSY breaking~\cite{Buchmuller:2005rt,Buchmuller:2006nx}.}.
\end{itemize}
The outline of the paper is the following: 
in the next section we present the most relevant features of TGM and set the framework for the cosmological
analysis. In the presence of exact R-parity the outcome is that 
for a typical TGM spectrum the DM relic density is overabundant and the decay of he NLSP is in tension with the BBN constraints.   
On the other hand a small amount of R-Parity Violation (RPV) can easily restore the agreement with the BBN and, at the same time, reduces the amount of gravitinos produced by the NLSP decay.
Hence we study how RPV can be obtained in the context of an $SO(10)$ model of TGM and we show that 
the GUT structure of the theory motivates the restriction to bilinear RPV. We finally analyze the cosmology 
of such a scenario and identify the phenomenologically viable regions of the parameter space.

\section{$SO(10)$ Tree Level Gauge Mediation}
\label{so10tgm}

As already mentioned in the introduction we are going to analyze the $SO(10)$ TGM model presented in~\cite{Nardecchia:2009ew}. 
We provide here a brief overview of the main features of the low-energy TGM spectrum. 

The $U(1)_X$ gauge group responsible for the TGM mechanism is identified with 
the abelian factor external to $SU(5)$ in the embedding $SU(5) \otimes U(1)_X \supset SO(10)$.
After the one-step breaking $SO(10) \to G_{SM}$ at the scale $M_{G}$, 
all the effects of TGM and the GUT physics are encoded in the MSSM boundary conditions at the GUT scale. 

The ratios among the tree level soft masses $(\tilde m^2_Q)_\text{tree}$ depend only on the embedding
of the MSSM chiral superfield $Q$ into the $SO(10)$ representations. 
It is useful to consider the decomposition of the $10$ and $16$ of $SO(10)$ with respect to the subgroup $SU(5) \otimes U(1)_X$, namely
\begin{equation}
\label{SO10repdec}
16 = 10_1 \oplus \overline{5}_{-3} \oplus 1_5 \, , \qquad 10 = 5_{-2} \oplus \overline{5}_{2} \, ,
\end{equation}
while the decomposition of the $\overline{16}$ follows from that of the $16$.
In order to avoid negative soft terms contributions (cf.~\eq{soft1}), all the MSSM matter superfields ($q,u^c,e^c,d^c,\ell$) must have same sign under $U(1)_X$.
This condition is fulfilled if
\begin{equation}
\label{pembedd}
q \oplus u^c \oplus e^c = 10_1 \subset 16 \, , \qquad  d^c \oplus \ell = \overline{5}_2 \subset 10 \, .
\end{equation} 
The MSSM Higgses are embedded in linear combinations of 10, 16 and $\overline{16}$. 
According to this discussion the TGM contribution to the bilinear soft masses is 
\begin{equation}
\label{eq:massrelations}
(\tilde m^2_Q)_\text{tree}=
\left\{
\begin{array}{l}
2 \, \tilde{m}^2_{10} \qquad Q=d^c,\ell \\
\tilde{m}_{10}^2 \quad\quad\ \ Q=q,u^c,e^c \\
-2 \, \tilde{m}^2_{10} < (\tilde{m}^2_{h_u})_\text{tree} < 3 \, \tilde{m}^2_{10}  \\
-3 \, \tilde{m}^2_{10} < (\tilde{m}^2_{h_d})_\text{tree} < 2 \, \tilde{m}^2_{10} 
\end{array}
\right. \, ,
\end{equation}
where $\tilde{m}^2_{10}$ is a universal mass parameter.

Gaugino masses are generated at the one-loop level as in standard gauge mediation. We call $M_{1/2}$ the common gaugino mass at $M_{G}$\footnote{We are not taking into account here the effect of having a hierarchical messenger spectrum, an analysis of this scenario will be presented in Ref.~\cite{monaco}.}.

The mechanism responsible for gaugino masses unavoidably generates a two-loop level contribution to the sfermion masses. 
The final expression at the GUT scale is given by
\begin{equation}
\label{eq:2loop}
\tilde m^2_Q = (\tilde m^2_Q)_\text{tree} + 2\, \eta \, C_Q M^2_{1/2} \, ,
\end{equation}
where $C_Q$ is the total SM quadratic Casimir 
\begin{equation}
\label{casimirs}
\begin{tabular}{c|ccccccc}
$Q$ & $q$ & $u^c$ & $d^c$ & $\ell$ & $e^c$ & $h_u$ & $h_d$ \\
\hline
$C_Q$ & 21/10 & 8/5 & 7/5 & 9/10 & 3/5 & 9/10 & 9/10
\end{tabular}  \, .
\end{equation}
The parameter $\eta > 0$ gives the relations between the two-loop contribution to the sfermion masses and that to the gaugino masses 
squared\footnote{In the notation of~\cite{Giudice:1998bp} this parameter can be identified with the ratio $\eta=\Lambda^2_S / \Lambda^2_G$.}. 
The precise value of $\eta$ depends on the details of the messenger sector. For instance in standard (loop) gauge mediation with one messenger chiral superfield 
this parameter is precisely $\eta= 1 / n$, where $n$ is the Dynkin index of the vector-like pair of messengers. In most of our analysis we will set $\eta=1$. 

In order to clearly discriminate TGM as the mechanism responsible for the soft scalar masses  
we will focus the attention on the regions of the parameter space where the values of $\tilde{m}_{10}$ and $M_{1/2}$ are such that 
TGM is responsible for the leading contribution to the sfermion masses. 
We define the dominance of TGM by requiring that TGM contributes to the low energy value of each sfermion mass by an amount of at least $50 \%$. 
Including also the running effects (cf.~\app{ap:1looprge} for further details) this translates into the condition
\be
\label{eq:tgmdominance}
\tilde{m}_{10} \gtrsim \left( 5.2 + 4.2 \,  \eta \right)^{1/2} M_{1/2} \, ,
\ee
which, for $\eta=1$, reduces to $\tilde{m}_{10} \gtrsim 3.1 \, M_{1/2}$.

On top of $\tilde{m}_{10}$ and $M_{1/2}$ the other MSSM parameters relevant at low-energy are $\tan\beta$, $\mu$, $B\mu$ and the $A$-terms.
Relating the $\mu$-term to SUSY breaking is a model-dependent issue\footnote{We just mention that TGM offers a new solution for the $\mu$-problem,
where the $\mu$ is also responsible for triggering the SUSY breaking~\cite{Nardecchia:2009nh}.}. 
Here we will just fix $\mu$ and $B\mu$ in such a way that they satisfy the electroweak symmetry breaking conditions. 
In addition we assume $\mu > 0$.
The $A$-terms are set to zero at the GUT scale. 
In general, since they do not arise at the tree level, 
they are expected to be smaller than the bilinear soft masses. 

In the case in which SUSY is broken only by the F-term responsible for sfermion masses,  the gravitino mass is directly related to $\tilde{m}_{10}$ by the relation~\cite{Nardecchia:2009ew}
\begin{equation}
\label{eq:gravitinomass}
{m}_{3/2}\approx 15 \,  \mbox{GeV}\left(\frac{{\tilde{m}}_{10}}{1 \ \mbox{TeV}}\right) \, .
\end{equation}
We stress again, due its cosmological relevance, that the magnitude of the gravitino mass is a peculiar prediction of TGM and its embedding into a GUT.

Let us close this section with a couple of comments regarding the spectrum. 
The first one is about the nature of the NLSP, being a cosmologically relevant issue. 
It turns out that in most of the parameter space the NLSP is a Bino-like neutralino. 
This is easily understood because, according to~\eq{eq:tgmdominance}, 
scalars are expected to be heavier than gauginos. 
Moreover, the running of gaugino masses (cf.~\eq{eq:1loopgaugino}) is such that $M_1 < M_2$.
The second comment regards the viability of the model in the light of the recent LHC exclusion limits~\cite{Aad:2011ib}. 
The condition in~\eq{eq:tgmdominance} can be translated (cf.~\eqs{mq}{muc} in~\app{ap:1looprge}) in terms of the values of the physical stop mass 
$\tilde{m}_t$ and the gluino one $m_{\tilde{g}}$, yielding
\begin{equation}
\tilde{m}_t \gtrsim 1.2 \ \textrm {TeV}  \,  \left( \frac{m_{\tilde{g}}}{700 \ \textrm {GeV} } \right) \, ,
\end{equation}
which shows that a sizable part of the parameters space is still viable and testable at the LHC.

\section{Gravitino Dark Matter with R-parity}
\label{relicdensityandBBN}

In this section we investigate the cosmological aspects of TGM in the R-parity conserving case. 
The cosmology is deeply influenced by the behavior of the NLSP. 
Indeed, due to the presence of R-parity, it affects the DM relic density by decaying into gravitinos. 
Being the rate of such a decay Planck-suppressed, gravitinos will be produced after the freeze-out of the NLSP and 
potentially also after the onset of the BBN. 
We then refer to these gravitinos as non-thermal and their abundance is given by
\begin{equation}
\label{eq:Grd}
{\Omega}_{3/2}^{NT}h^2=\frac{{m}_{3/2}}{{m}_{NLSP}}{\Omega}_{NLSP} \, h^2 \simeq 3 \times 10^{-2} \frac{\tilde{m}_{10}}{M_{1/2}}{\Omega}_{NLSP} \, h^2 \, ,
\end{equation} 
where $\Omega_{NLSP}$ is the expected NLSP relic density as if it were stable.
  
${\Omega}_{3/2}^{NT}$ has to be added to the abundance ${\Omega}_{DM}^T$ of gravitinos produced by thermal processes in the early Universe, 
in order to match the WMAP-7 value $\Omega_{DM}h^2=0.1123 \pm 0.0035$~\cite{Komatsu:2010fb}. 
The thermal component can be computed once the rates of the relevant processes are known as a function of the MSSM parameters 
(see e.g.~\cite{Rychkov:2007uq}) and it depends on the reheating temperature $T_{RH}$, i.e.~the temperature which sets the beginning of the radiation domination era. 
 
\begin{figure}[htbp]
\subfloat{\includegraphics[width=7.4 cm, height= 7.4 cm, angle=360]{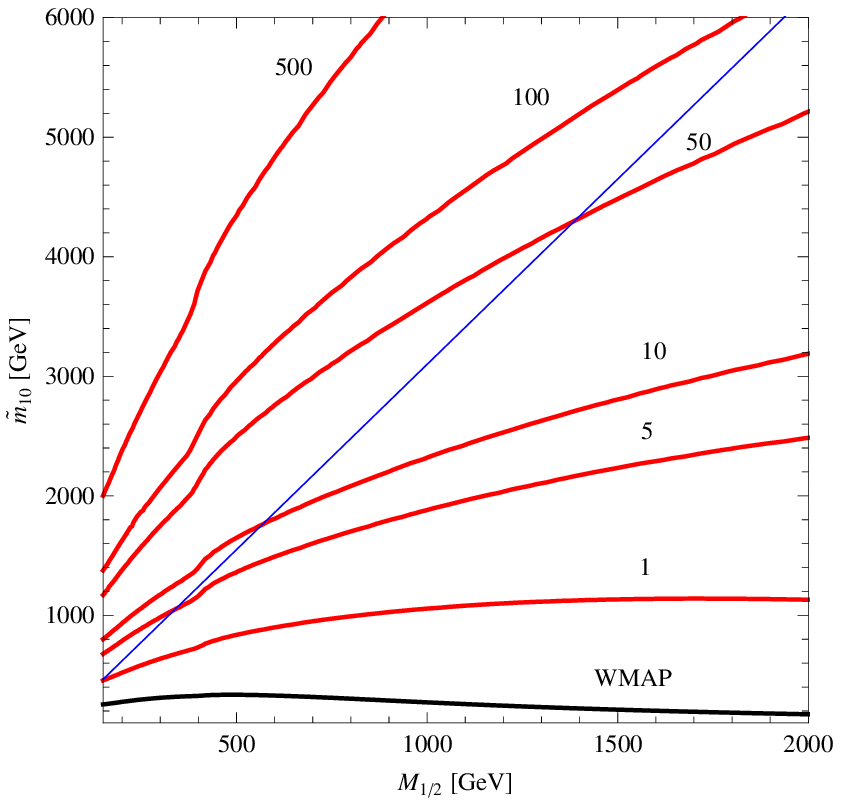}}
\subfloat{\includegraphics[width=7.4 cm, height= 7.4 cm, angle=360]{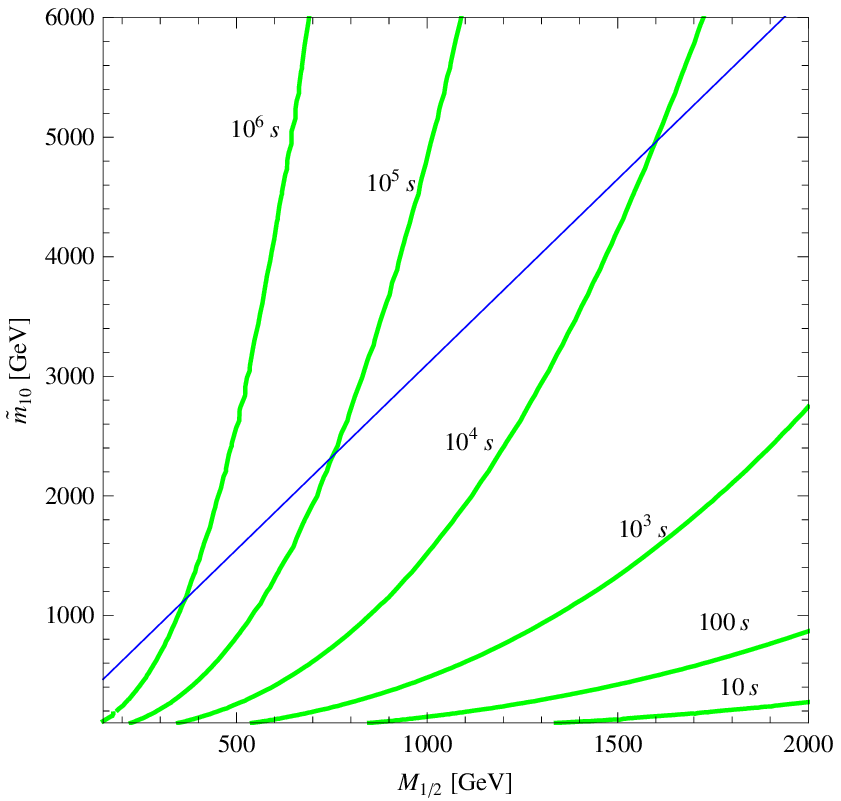}}
\caption{
Contours of the non-thermal component of the gravitino relic density (left panel) and of the neutralino lifetime (right panel) 
 in the plane $(\tilde{m}_{10}, M_{1/2})$, 
with the other MSSM parameters fixed as explicitly said in the text. 
The black line represents the non-thermal relic density fit of the WMAP-7 value.
The blue line represents the relation $\tilde{m}_{10}=3.1 \, M_{1/2}$ which sets the TGM dominance.}  
\label{fig:samples}
\end{figure}

The non-thermal component for the model in consideration is obtained by means of the scaling formula in~\eq{eq:Grd} with 
${\Omega}_{NLSP}$ computed through the numerical code DARKSUSY~\cite{Gondolo:2004sc}. 
The results are reported in the left panel of~\fig{fig:samples}.
The outcome is that the scenarios in which TGM is the dominant mechanism for the generation of sfermion masses 
(cf.~the region in~\fig{fig:samples} above the blue line) are characterized 
by an overabundance of non-thermal gravitinos. 

This behaviour can be easily understood from the fact that Binos annihilate mainly into fermion pairs with a p-wave suppressed cross-section 
$\propto m^2_{\chi^0_1} / \tilde{m}_Q^4$, where $m_{\chi^0_1}$ is the lightest neutralino mass.
In such a case the NLSP relic density can be estimated by
\be
\label{eq:binoestimate}
\Omega_{NLSP} \, h^2 \approx 0.02 \times 10^3 {\left(\frac{m_{{\chi}_1^0}}{150 \, \mbox{GeV}}\right)}^{-2} {\left(\frac{\tilde{m}_Q}{1 \, \mbox{TeV}}\right)}^4 \, .
\ee
Combining this result with~\eq{eq:Grd} it is evident that the cosmological value of the DM relic density is largely 
overcome for $\tilde{m}_{10} > M_{1/2}$, as predicted by TGM. 

The decay of the NLSP is also responsible for another important cosmological issue.  
Indeed the gravitino injection is accompanied with the production of SM particles which can trigger either electromagnetic or hadronic showers. 
In turn these may upset the BBN calculations of the light-element abundances.  
As shown in the right panel of~\fig{fig:samples} the lifetime of the neutralino is much bigger than the time of the onset of the BBN
also for moderate values of the ratio $\tilde{m}_{10} / M_{1/2}$. 

We can easily understand this behaviour from the functional dependence of the decay rates. A Bino NLSP mainly decays either into a $Z$ boson and a gravitino or into a photon and a gravitino and the relevant rates can be expressed as~\cite{Feng:2004mt}
\be
 \label{eq:nzg}
\Gamma({\chi}_{1}^{0} \rightarrow Z \tilde{G}) = \frac{{\sin^{2}{\theta}_{W}}}{48\pi {M}_{P}^{2}}\frac{{m}_{{\chi}^{0}_{1}}^{5}}{{m}_{3/2}^{2}} \, , 
\qquad \Gamma({\chi}_{1}^{0}\rightarrow \gamma \, \tilde{G})= \frac{{\cos^{2}{\theta}_{W}}}{48\pi {M}_{P}^{2}}\frac{{m}_{{\chi}^{0}_{1}}^{5}}{{m}_{3/2}^{2}} \, .
\ee
In particular the NLSP can induce hadronic showers from the decay of the Z boson. These have the deepest impact on the BBN and   
hence suffer from the most severe bounds. The constraints are both on the lifetime and the abundance of the NLSP, and
they basically exclude all neutralinos with a lifetime greater than $10^{-2}$ s~\cite{Covi:2009bk}. 
We conclude that is not possible to obtain a viable cosmology in the TGM setup with R-parity conservation. 

It is also evident that the BBN bounds, together with the issue of the overabundance of gravitinos, 
can be evaded in the presence of some mechanism which suppresses the NLSP abundance. In this paper we will consider the case of a small amount of R-parity violation\footnote{An alternative scenario could be the dilution of the NLSP abundance due to the entropy released by the decay of some heavy state before the onset of BBN~\cite{Hasenkamp:2010if, Arcadi:2011ev}.}.
 
\begin{figure}[htbp]
\subfloat{\includegraphics[width=7.4 cm, height= 7.4 cm, angle=360]{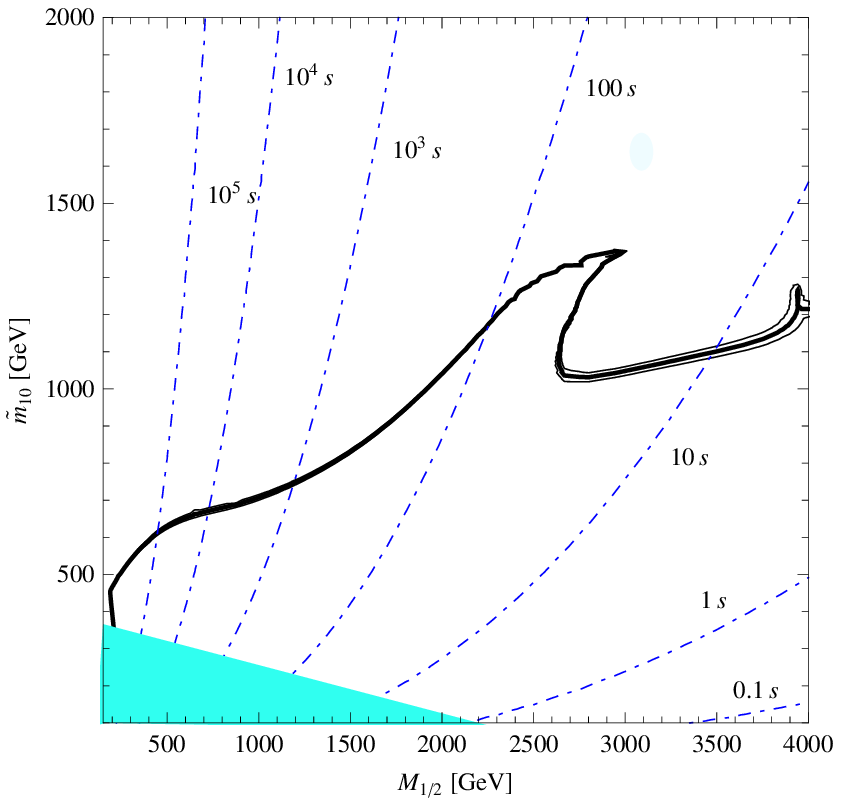}}
\subfloat{\includegraphics[width=7.4 cm, height= 7.4 cm, angle=360]{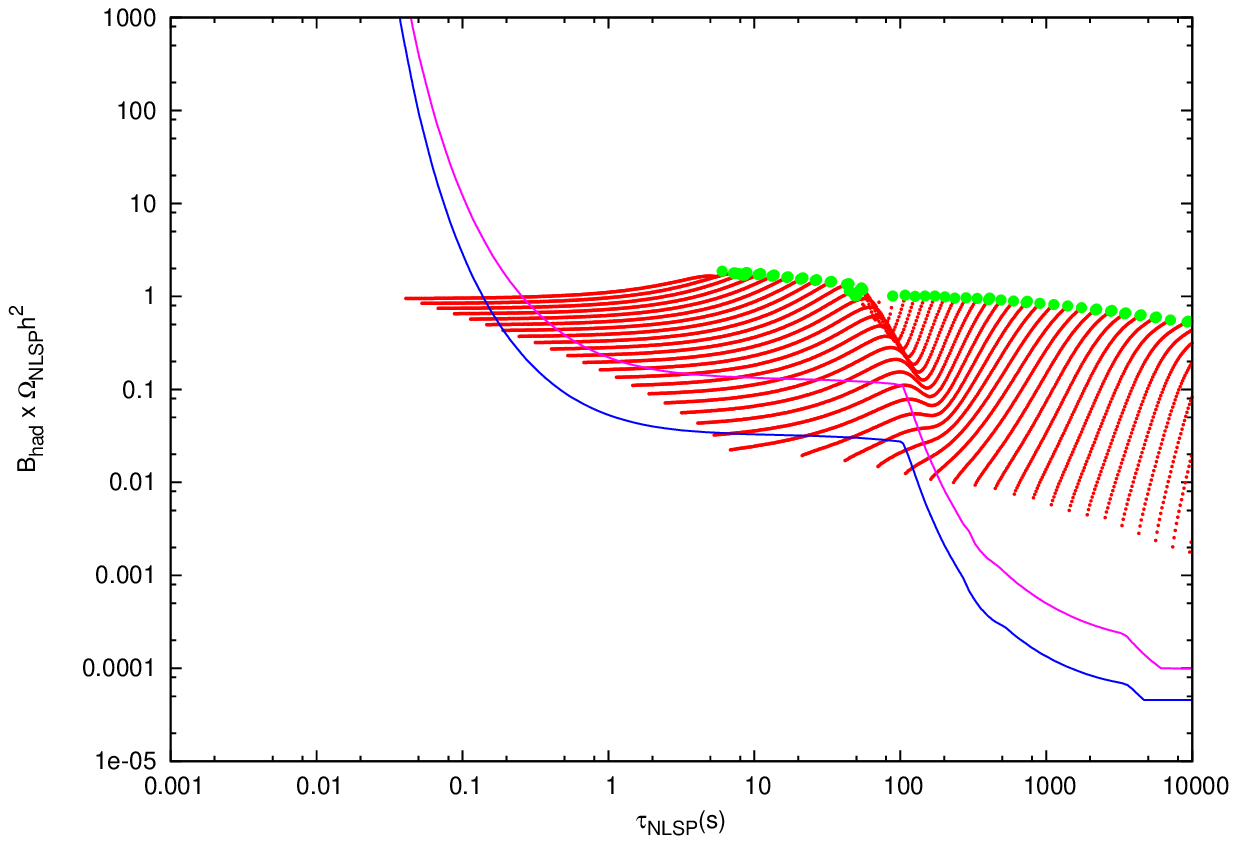}}
\caption{Left panel: WMAP-7 value of the non thermal contribution to the DM relic density, within a 1-$\sigma$ deviation. 
Blue dot-dashed lines represent the contours of the NLSP lifetime. 
Right panel: points of the left panel plot which give a non-thermal gravitino relic density lower (red points) or within 1-$\sigma$ (green points) 
with respect to the WMAP-7 expectations. The blue and the violet lines represent, respectively, the BBN bound for a 100 GeV and 1 TeV decaying particle \cite{Jedamzik:2006xz,Covi:2009bk}.}
\label{fig:ben2}
\end{figure}

For completeness we mention that, by relaxing the condition~\eq{eq:tgmdominance}, it is possible to realize scenarios with a viable cosmology. 
Here we report two examples.
The first one is given by the enhancement of the s-channel annihilation of neutralinos into a bottom pair mediated by the CP-odd Higgs, typically occuring at high values of $\tan\beta$. This enhancement is particularly strong in the region ${m}_{A} \sim 2 {m}_{\chi}$, where the annihilation cross section can also 
become resonant. 
As shown in~\fig{fig:ben2}, this scenario is realized only if one deviates sensibly from the relation~\eq{eq:tgmdominance} and hence in a region where standard (loop) gauge mediation is the dominant mechanism for the sfermion mass generation.   

The second example is the case in which the NLSP is a stau. 
This is possible for high values of $\tan\beta$ and $\eta < 1$. In this case the negative Yukawa corrections to the third family sfermions, originating at high $\tan\beta$, can drive the mass of the lightest stau below that of the lightest neutralino.
 The stau mainly decays into a gravitino and a tau lepton with the following rate~\cite{Feng:2004mt}
\begin{equation}
\label{eq:staurate}
\Gamma\left(\tilde{\tau} \rightarrow \tau\,\tilde{G}\right)=\frac{1}{48 \pi M_P^2} \frac{m_{\tilde{\tau}}^5}{m_{3/2}^2}
\, ,
\end{equation}
without producing hadronic showers. 
In this case the strongest BBN bounds are given by the formation of bound states with Helium nuclei which can alter the primordial
abundance of Lithium. This process is referred as catalyzed BBN \cite{Pradler:2007is,Pospelov:2008ta} and implies an upper bound of around 
$5 \times 10^3 \ \mbox{s}$ on the stau lifetime\footnote{It should be mentioned that in settings with substantial left-right mixing of the stau mass eigenstates 
the BBN bounds can be evaded even for higher lifetimes \cite{Ratz:2008qh}.}.  
Given the rate~\eq{eq:staurate} this translates into the requirement of a stau mass greater than $200 \div 300$ GeV.

\section{Gravitino Dark Matter without R-parity}

The outcome of~\sect{relicdensityandBBN} is that the configurations of TGM dominance, $\tilde{m}_{10} \gtrsim 3.1 \, M_{1/2}$, 
cannot fulfill the cosmological bounds regarding the DM relic density and the BBN. 
On the contrary these bounds can be evaded in presence of a small amount of R-parity breaking 
(see e.g.~Ref.~\cite{Buchmuller:2007ui}). Indeed in such a case the NLSP can decay only into SM particles before the onset of the BBN, without 
contributing to the DM relic density in the form of non-thermal gravitinos.    
At the same time the thermally produced gravitinos should be stable enough in order to reproduce the correct DM relic density.  
Then, since the NLSP lifetime does not depend anymore on the gravitino mass, 
a hierarchy between sfermion and gaugino masses, 
as expected in TGM, is in principle achievable. 

In the following we will present and analyze an $SO(10)$ TGM model with explicit R-parity violation. 
In particular we will show that applying the bounds from proton decay the GUT structure of the theory guarantees the suppression 
of all the trilinear RPV couplings. This motivates the restriction to only bilinear RPV operators in the cosmological analysis.

\subsection{An R-parity violating $SO(10)$ model}
\label{RPVinTGM}

The R-parity is a $Z_2$ symmetry which distinguishes the SM fields from their super-partners.
By assigning a positive charge to the former and negative one to the latter it provides for instance the stability of the LSP. 
It is useful to rephrase the R-parity in a slightly different language~\cite{Martin:1992mq} 
\be
\label{defrp}
R_p = (-)^{3(B-L)+2s} \, ,
\ee 
where $s$ is the spin 
quantum number. 
As long as the angular momentum is conserved, the R-parity is essentially equivalent 
to a $Z_2$ Matter (M)-parity, defined as
\be
\label{defmp}
M_p = (-)^{3(B-L)} \, .
\ee
\eq{defmp} suggests that in theories in which $B-L$ is gauged the M-parity 
can be viewed as a discrete subgroup of a local $U(1)_{B-L}$, thus providing a possible link between the amount of 
R-parity violation and the $B-L$ breaking scale. 
However, though one of the $SO(10)$ generators can be identified with $B-L$ 
when acting on the spinorial representation $16_F$,
this does not apply to TGM because of the peculiar embedding of the SM fermions 
into $16_F \oplus 10_F$ (cf.~\eq{pembedd}).

Thus the $SO(10)$ gauge symmetry does not protect against the appearance of R-parity violating interactions and the simplest way 
to forbid them is to impose an M-parity \emph{external} to $SO(10)$ which distinguishes the matter superfields (with negative charge) from the Higgs ones 
(carrying positive charge)~\cite{Nardecchia:2009ew}. 

Our approach will be that of relaxing the presence of this extra M-parity and, by considering all the operators compatible with the $SO(10)$ symmetry, 
we will prove the existence of R-parity violation in the low-energy effective theory.
In addition we will assume that the R-parity violating operators are suppressed when 
compared to their R-parity conserving counterparts, in order to avoid an unacceptable amount of lepton and baryon number violation. 
Addressing the issue of the origin of such a small amount of R-parity breaking is anyway beyond the scopes of this work.

For definiteness let us focus on a supersymmetric $SO(10)$ model featuring the following minimal set of Higgs 
representations: 
$54_H \oplus 45_H \oplus 16_H \oplus \overline{16}_H \oplus 10_H$\footnote{To be complete one should also 
add a $16'_H \oplus \overline{16}'_H$ representation which is responsible for the SUSY breaking~\cite{Nardecchia:2009ew}. 
For simplicity we will carry on our analysis in the supersymmetric limit, assuming that the conclusions 
regarding the gauge symmetry breaking and the fermion mass spectrum are only marginally affected by the 
SUSY breaking sector.}. 
As shown in~\app{SO10model} this field content is sufficient in order to break $SO(10)$ 
down to the SM at the renormalizable level (cf.~\app{SO10modelsb}) 
and to give mass to the SM fermions (cf.~\app{SO10modelyu}).  
As already mentioned the MSSM matter superfields span over three copies of $16_F \oplus 10_F$ 
in such a way that they are embedded in the $SU(5)$ representations $10 \supset 16_F$ and $\overline{5} \supset 10_F$ 
(cf.~\eq{pembedd}).  
The conditions to be fulfilled in order to obtain such a ``pure'' embedding are detailed in~\app{SO10modelyu}.

The superpotential can be schematically written as
\be
W = W_H + W_Y + \delta W_{RPV} \, ,
\ee
where $W_H$ and $W_Y$ are the Higgs and the Yukawa components  
\begin{align}
\label{WHiggs}
W_H &= \left( \mu_{54} + \eta_{54} 45_H + \lambda_{54} 54_H \right) 54_H^2
+ \mu_{45} 45_H^2
+ \left( \mu_{10} + \lambda_{10} 54_H \right) 10_H^2 \nn \\
& + \left( \mu_{16} + \lambda_{16} 45_H \right) 16_H \overline{16}_H 
+ \lambda_{16-10} 16_H^2 10_H + \overline{\lambda}_{16-10} \overline{16}_H^2 10_H \, , \\
\label{WYuk}
W_Y &=  Y_{10}^{ij} 16_F^i 16_F^j 10_H + Y_{16}^{ij} 16_F^i 10_F^j 16_H + \left( M_{10}^{ij} + \eta^{ij} 45_H + \lambda^{ij} 54_H \right) 10_F^i 10_F^j \, , 
\end{align}
while $\delta W_{RPV}$ is the R-parity violating piece
\begin{multline}
\label{WRPV}
\delta W_{RPV} = \left( \tilde{\mu}_{10}^i + \tilde{\eta}_{10}^i 45_H + \tilde{\lambda}_{10}^i 54_H \right) 10_F^i 10_H 
+ \left( \tilde{\mu}_{16}^i  + \tilde{\lambda}_{16}^i  45_H \right) 16_F^i \overline{16}_H \\
+ \tilde{\rho}^i 16_F^i 16_H 10_H 
+ \tilde{\sigma}^i 10_F^i 16_H 16_H + \tilde{\overline{\sigma}}^i 10_F^i \overline{16}_H \overline{16}_H
+ \tilde{\Lambda}^{ijk} 16_F^i 16_F^j 10_F^k \, .
\end{multline}
Notice that without M-parity the separation between the $F$ and the $H$ superfields is somehow artificial.
However, since we consider $\delta W_{RPV}$ as a perturbation, we can still 
retain $W_H$ responsible for the symmetry breaking and $W_Y$ for the (charged fermions) Yukawa sector. 
On the other hand the situation about neutrino masses is subtler, being RPV potentially responsible for sizable contributions to them. 
We will comment later on the generation of neutrino masses in our model.

In~\app{SO10modelrp} we provide  
an existence proof of the R-parity violating operators in the MSSM effective 
theory. In order to obtain the low-energy superpotential one has to project the operators of~\eq{WRPV} 
on the representations containing the MSSM fields. 
Here we just report the results of this operation leaving most of the technical details in~\app{SO10modelrp}. 

Bilinear $R$-parity violation in the effective superpotential is induceded by operators containing just one $F$ superfield in~\eq{WRPV}, leading to
\be
\label{eq:WR}
W^{eff}_{RPV} \supset \mu_i \, \ell_i h_{u} \, ,
\ee
where the expression of $\mu_i$ in terms of the original couplings is given in~\eq{muiGUT}.

Notice that in the ``pure'' matter embedding of TGM only some of the operators have projections on the MSSM fields. In particular 
the trilinear operator relative to the coupling $\tilde{\Lambda}^{ijk}$ does not contribute to the effective theory.
On the other hand the phenomenological viability of the model, within the minimal choice of representations at hand, requires the presence
of non-renormalizable operators (cf.~again~\app{SO10model}). By relaxing renormalizability there is an additional source of R-parity violation 
given by the operator
\be
\label{WRPVNR}
\frac{\tilde{\Lambda}_{ijk}^{NR}}{M_P} \, 10^i_F 10^j_F 16^k_F \vev{\overline{16}_H} 
\supset 
\lambda_{ijk}^{NR} \, \ell_i \ell_j e^{c}_k + \lambda'^{NR}_{ijk} \, d^{c}_i \ell_j q_k + \lambda''^{NR}_{ijk} \, d^{c}_i d^{c}_j u^{c}_k \, ,
\ee
where 
\be
\label{SU5tril}
\lambda_{ijk}^{NR} = \frac{1}{2} \lambda'^{NR}_{ijk} = \lambda''^{NR}_{ijk} = 
\frac{\tilde{\Lambda}_{ijk}^{NR} \, V^{16}}{M_P} \equiv \Lambda_{ijk} \, .
\ee
Notice that the relation in~\eq{SU5tril} gives a correlation between the baryon ($\lambda''$) and lepton number ($\lambda$, $\lambda'$) 
violating couplings.

We should also mention that $\lambda''$ receives an additional contribution when combining the bilinear operators in~\eq{WRPV} 
with the Yukawa ones. This is obtained by projecting the Higgs fields on the heavy triplet components and integrating them out. 
This last contribution, labeled $\lambda''_{T}$, is shown explicitly in~\eq{lambdaGUT}.

In the end the structure of the induced superpotential in the MSSM effective theory is given by:
\be
W^{eff}_{RPV}=  \mu_i \, \ell_i h_{u}  + 
\lambda_{ijk} \, \ell_i \ell_j e^{c}_k + \lambda'_{ijk} \, d^{c}_i \ell_j q_k + \lambda''_{ijk} \, d^{c}_i d^{c}_j u^{c}_k \, ,
\ee
where 
\be
\label{trilrel}
\lambda = \Lambda \, , \qquad \lambda' = 2 \Lambda \, , \qquad \lambda'' = \Lambda + \lambda''_{T} \, .
\ee
The strongest constraints on the R-parity violating interactions are due to proton decay. 
In particular this enforces severe bounds on the products of couplings 
$\lambda \lambda''$, $\lambda' \lambda''$ (cf.~e.g.~\cite{Barbier:2004ez} for an exhaustive list). 
For kinematical reasons the most stringent ones apply to products involving dominantly the first two light generation indices. 
For instance the process $p \rightarrow {\pi}^{0}{e}^{+}$ severely constrains the product 
\begin{equation}
\label{eq:proton_1}
\lambda_{k11}^{\prime} \lambda^{\prime \prime}_{k11} \lesssim {10}^{-26} \left( \frac{\tilde{m}}{1 \ \text{TeV}} \right)^2 \, ,
\end{equation}
with $k=2,3$ and $\tilde{m}$ being the sfermion mass scale.   
Though the extension of the analysis at the one-loop level sets weaker bounds for the couplings relative to the 
second and third generation~\cite{Smirnov:1996bg}, in our setup the structure of the trilinears is constrained by the GUT symmetry 
which implies much stronger bounds with respect to the general case. The most conservative 
bound on all the R-parity violating trilinear couplings in the presence of a GUT relation such as that in~\eq{SU5tril} is given by~\cite{Smirnov:1995ey}
\be
\label{Lambdabound}
\Lambda \lesssim {10}^{-10} \left( \frac{\tilde{m}}{1 \ \text{TeV}} \right)^2. 
\ee  
Barring extremely accurate cancellations between the two unrelated components $\Lambda$ and $\lambda_T$ in the expression for $\lambda''$ 
(cf.~\eq{trilrel}), the bound in~\eq{Lambdabound} is automatically translated onto $\lambda$ and $\lambda'$.

The bounds on the trilinear RPV couplings just derived are very strong, making them harmless for the cosmological analysis.
In light of this result, a motivated setup for the cosmological analysis is bilinear R-parity violation. 
Then the R-parity violating superpotential simply reads
\be
\label{eq:star_superpotential}
 W^{eff}_{RPV} = \mu_i \, \ell_i h_u  \, .
 \ee 
In the effective theory one also expects R-parity violating couplings in the soft scalar potential,
depending on the details of the SUSY breaking sector. 
As we will show in the next section, the main cosmological constraints apply to the bilinear soft terms
\be
\label{eq:softpotential}
V^{\rm soft}_{RPV}= B_i \, \tilde{\ell}_i h_u + \tilde{m}^2_{h_d \ell_i} h^{\dagger}_d \tilde{\ell}_i + \textrm{h.c.} \, .
\ee

\subsection{Cosmological analysis}
\label{generalbounds}

Without R-parity, and thus lepton number conservation, there is no a priori distinction between ${h}_{d}$ and ${\ell}_{i}$. 
This is reflected by the fact that~\eqs{eq:star_superpotential}{eq:softpotential} induce a non-vanishing VEV for 
the sneutrino fields once electroweak symmetry is broken~\cite{Hall:1983id,Barbier:2004ez}. 

Since ${h}_{d}$ and ${\ell}_{i}$ have the same quantum numbers 
one can always operate a redefinition of these fields through a unitary transformation. 
Typically one can use this transformation in order 
to rotate away either the sneutrino VEVs $v_i$ or the bilinear couplings $\mu_i$. 

Since we consider~\eq{eq:star_superpotential} as a perturbation of the R-parity conserving theory, 
implying in particular $\mu_i \ll \mu$, it is convenient to define the following linear transformation on the superfields
\be
\label{primedbasis}
h_d \rightarrow \hat{h}_{d}={h}_{d}+{\epsilon}_{i}{\ell}_{i} \, , \qquad
\ell_i \rightarrow \hat{\ell}_{i}={\ell}_{i}-{\epsilon}_{i}{h}_{d} \, ,
\ee
with $\epsilon_i=\mu_i / \mu$, which rotates away the bilinear term in~\eq{eq:star_superpotential} up to $\mathcal{O}(\epsilon_i^2)$ corrections. 
The expressions of the R-parity violating couplings in the hatted basis 
can be found for instance in Ref.~\cite{Barbier:2004ez}. 
We just point out that the transformation in~\eq{primedbasis} induces trilinear lepton violating couplings $\hat{\lambda}_{ijk}$, $\hat{\lambda}'_{ijk}$ of the form
\begin{equation}
\label{eq:eff_lambdas}
\hat{\lambda}_{ijk}=-{(Y_e)}_{ik} \epsilon_j+{(Y_e)}_{jk} \epsilon_i \, , \qquad 
\hat{\lambda}'_{ijk}=-{(Y_d)}_{ik} \epsilon_j \, , \qquad 
\end{equation}
where $Y_e$ and $Y_d$ represent, respectively, the SM Yukawas of the charged-leptons and the down-quarks.
In the hatted basis the VEVs of the sneutrinos are given by~\cite{Buchmuller:2007ui}
\be
\label{eq:vi}
v_i
\equiv -{\xi}_{i}\langle {h}_{d}\rangle=
-\frac{\hat{B}_{i}\tan\beta+\hat{\tilde{m}}_{h_d \ell_i}^{2}}{\hat{\tilde{m}}^{2}_{\ell_i}+\frac{1}{2}{M}_{Z}^{2}\cos2\beta}\langle {h}_{d}\rangle \, .
\ee
Then we can express $\xi_i$ in terms of the original parameters obtaining, at the leading order in $\epsilon_i$, 
\be
\label{eq:xi}
\xi_i \approx \frac{(B_i - \epsilon_i B) \tan\beta + \tilde{m}_{h_d \ell_i}^2 + \epsilon_i (\tilde{m}^{2}_{\ell_i} - \tilde{m}^2_{h_d}) }
{\tilde{m}^{2}_{\ell_i}+\frac{1}{2}{M}_{Z}^{2}\cos2\beta} \, .
\ee
Given the model dependence of the soft terms $B_i$ and $\tilde{m}_{h_d \ell_i}$, RPV is described at low-energy by the  
six parameters $\xi_i$ and $\epsilon_i$. 
On the other hand, by inspecting ~\eq{eq:xi} and~\eq{eq:massrelations} it turns out that, barring cancellations, 
the parameters $\xi_i$ are at least of the order of $\epsilon_i$. 

In what follows we will analyze the bilinear RPV model introduced in the previous section and identify the range of viability of the 
R-parity violating couplings. 
The analysis is organized according to the following points:
\begin{itemize}
\item NLSP lifetime and BBN.
\item Gravitino lifetime and cosmic rays.
\item Neutrino masses through RPV.
\item Gravitino relic density and thermal leptogenesis.
\end{itemize}

\subsubsection*{NLSP lifetime and BBN}
 
The main motivation for the introduction of R-parity violation is to restore the agreement 
between the decay of the NLSP and BBN.  
In our setup the most efficient processes are induced by the sneutrino VEVs which 
mixes the $Z$ ($W$) boson with a neutrino (charged-lepton) and a neutralino \cite{Roy:1996bua,Mukhopadhyaya:1998xj}.

A Bino can decay into a $W$ boson and a charged-lepton or into a $Z$ boson and a neutrino with the typical rates~\cite{Bobrovskyi:2010ps}
\begin{align}
\label{Zto}
& \Gamma\left(\chi_1^0 \rightarrow Z \,\nu\right)
= \frac{{G}_{F} \, {m}_{\chi_1^0}^{3}}{4\pi\sqrt{2}} \frac{{\sin}^2 \theta_W {\cos}^2 \beta}{M_1^2}\xi^2 \, , \\
\label{Wto}
& \Gamma\left(\chi_1^0 \rightarrow {W}^{\pm}{l}^{\mp}\right)
= \frac{{G}_{F} \, {m}_{{\chi_1^0}}^{3}}{2\pi \sqrt{2}} \frac{{\sin}^2 \theta_W {\cos}^2 \beta}{M_1^2}\xi^2 \, ,
\end{align}
where $\xi \equiv \sqrt{\xi_1^2 + \xi_2^2 + \xi_3^2}$.
The NLSP may also decay into three fermions by means of the couplings $\hat{\lambda}$ and $\hat{\lambda}'$ in~\eq{eq:eff_lambdas}
with a typical rate of the form
\begin{equation}
\label{Grate}
\Gamma_{\rm 3-body}=\frac{g^2 {| \hat{\lambda}^{\prime} |}^2}{1024 \, \pi^3} \frac{m_{\chi^0_1}^5}{\tilde{m}^4_Q} \, .
\end{equation}
The same expression, divided by a factor of three, holds for the rates involving the coupling $\hat{\lambda}$. 
However the 3-body processes are highly suppressed with respect to the 2-body decay
\be
\label{eq:Br32}
BR(3-\mbox{body}/2-\mbox{body}) \approx 0.9 \times 10^{-5} {\left(\frac{\epsilon}{\xi}\right)}^{2} {\left(\frac{\tan\beta}{10}\right)}^4 {\left(\frac{\tilde{m}_Q}{1 \ \mbox{TeV}}\right)}^{-4}
 {\left(\frac{m_{{\chi}_1^0}}{150 \ \mbox{GeV}}\right)}^{4} \, .
\ee
In writing this expression we used~\eq{eq:eff_lambdas} and assumed $\epsilon_i \sim \epsilon$ for $i=1,2,3$. 
Hence the 3-body processes can be neglected for a typical TGM spectrum, barring cancellations in the sneutrino VEVs. 
The NLSP lifetime is then determined by~\eqs{Zto}{Wto}
\begin{equation}
\label{eq:taunlsp2body}
{\tau}_{\rm NLSP, \, 2-body} \approx 0.02 \ \mbox{s} \ {\left(\frac{{m}_{{\chi}_1^0}}{150 \ \mbox{GeV}}\right)}^{-1}{\left(\frac{\tan\beta}{10}\right)}^{2}
{\left(\frac{\xi}{{10}^{-10}}\right)}^{-2} \, ,
\end{equation}
which satisfies the BBN bounds, as reproduced in the right panel of~\fig{fig:ben2} for $\tau_{\rm NLSP} \gtrsim 10^{-2} $ s, thus implying 
$\xi \gtrsim 10^{-(10 \div 11)}$ depending on $m_{{\chi}_1^0}$. This is actually a rather conservative bound; lower  values of $\xi$ are allowed depending on the NLSP abundance\footnote{
On the other hand the given bound takes also into account the fact that a population of NLSP, potentially dangerous for BBN, 
could survive despite the low branching ratio of 3-body decays. 
From~\eq{eq:Br32} we see that this scenario does not occur 
for low/moderate values of $\tan\beta$ (cf.~also the lines in the second panel of~\fig{fig:ben2}).
Eventually it could be necessary to assume an even more conservative limit $\xi > 10^{-9}$.}.

\subsubsection*{Gravitino lifetime and cosmic rays}

The amount of R-parity violation is also constrained from above since the DM is not stable anymore. 
As it will be evident from the expressions below the gravitino is stable on cosmological time-scales, 
being its decay rate doubly suppressed both by the R-parity violating couplings and the Planck mass. 
On the other hand the small portion of the decaying gravitinos is able to leave an imprint on the cosmic ray spectrum. 

In our setup the main decay channel of the gravitino is into a neutrino and a photon~\cite{Takayama:2000uz}
\begin{equation}
\Gamma (\tilde{G} \rightarrow \gamma \, \nu)=\frac{1}{32\pi}\frac{{\left(M_2-M_1\right)}^2}{M_1^2 M_2^2}
M_Z^2 \sin^2\theta_W \cos^2\theta_W \cos^2\beta \, \xi^2 \frac{m_{3/2}^3}{M^2_{P}} \, .
\end{equation}
Then the associated lifetime can be estimated 
by\footnote{The trilinear couplings $\hat{\lambda}$ and $\hat{\lambda}'$ yield a negligible contribution to the gravitino lifetime. 
Indeed the typical rates are~\cite{Moreau:2001sr} 
\begin{equation}
{\Gamma}_{3/2 \, , \rm 3-body} = \frac{{\bar{\lambda}}^2}{18432 \, \pi^3}\frac{m^7_{3/2}}{\tilde{m}^4_Q} \, , \qquad 
\bar{\lambda} = 3 \, \hat{\lambda}', \ \hat{\lambda}
\end{equation}
which lead to the following lifetime
\begin{equation}
\label{taug3body}
{\tau}_{3/2 \, , \rm 3-body} \approx 6.5 \times {10}^{37} \ \mbox{s} \ {\left(\frac{\epsilon}{10^{-4}}\right)}^{-2}{\left(\frac{m_{3/2}}{10 \ \mbox{GeV}}\right)}^{-7}
{\left(\frac{\tan\beta}{10}\right)}^{-2}{\left(\frac{\tilde{m}_Q}{1 \ \mbox{TeV}}\right)}^{4} \, .
\end{equation}
}
\begin{equation}
\label{taug2body}
\tau \simeq 7.3 \times 10^{28} \ \mbox{s} \ {\left(\frac{\tan\beta}{10}\right)}^2 {\left(\frac{M_{1/2}}{300 \ \mbox{GeV}}\right)}^2
{\left(\frac{m_{3/2}}{15 \ \mbox{GeV}}\right)}^{-3}{\left(\frac{\xi}{10^{-7}}\right)}^{-2} \, .
\end{equation}
This process is expected to leave an imprint on the cosmic gamma ray spectrum in the form of an approximately 
monochromatic line at an energy depending on the gravitino mass. 
For higher values of $M_{1/2}$ and $m_{3/2}$ the 3-body processes mediated by off-shell gauge bosons
become also important and eventually dominant~\cite{Choi:2010jt,Diaz:2011pc}, 
implying the presence of an additional continuos component in the gamma ray spectrum. 
This kind of signals have been the subject of dedicated searches 
performed by the Fermi Large Area Telescope (LAT) collaboration~\cite{Abdo:2010nc,Abdo:2010nz}. 
Since none of the expected excesses in the gamma ray spectrum have been 
detected so far one obtains a lower bound on the gravitino lifetime. 

According to Refs.~\cite{Choi:2010jt,Vertongen:2011mu} the lower bound on the gravitino lifetime is approximatively of 
$10^{27 \div 29}$ s for gravitino masses in the range $10 \div 80$ GeV and $M_{1/2}$ in the range
$100 \div 1000$ GeV. This translates into an upper bound for $\xi$ of about $\xi \lesssim 10^{-(6 \div 8)}$. 

For definiteness we mention that for the central values of $M_{1/2}$ and $m_{3/2}$ in \eq{taug2body}, which roughly correspond to the LHC bound quoted in \sect{so10tgm}, the limit on the gravitino lifetime is $10^{28}$ s which is satisfied for $\xi \lesssim 3\times 10^{-7}$. 

\subsubsection*{Neutrino masses trough RPV}
Summing up the results obtained until now, the outcome of the cosmological analysis is that 
the RPV coupling $\xi$ must lie in the window $10^{-11} < \xi < 10^{-6}$. R-parity violation can have, however, 
a wider impact on the phenomenology being a potential source of neutrino masses besides its cosmological 
role\footnote{In this respect it has been pointed out that the MSSM without R-parity could be a complete theory 
of the low-energy phenomena including neutrino masses, and it could even fit the cosmic ray anomalies of PAMELA and Fermi-LAT 
in terms of (decaying) gravitino DM~\cite{Bajc:2010qj}.}. 
Indeed the VEVs of the sneutrinos induce a mixing between neutrinos and neutralinos leading to a contribution to neutrino masses 
after integrating out the neutralinos~\cite{Hall:1983id}.
The result is a rank-one matrix hence giving mass to just one neutrino.
Explicitly one finds~\cite{Chun:1998ub}
\begin{equation}
\label{eq:mn3}
{m}_{{\nu}_{3}}={M}_{Z}^{2} \, {\xi}^{2}{\cos}^{2}\beta {\left(\frac{{M}_{1}{M}_{2}}{{M}_{1}{c}_{W}^{2}+{M}_{2}{s}_{W}^{2}}-\frac{{M}_{Z}^{2}}{\mu}\sin2\beta\right)}^{-1} \, .
\end{equation}
The complete neutrino spectrum can be then reconstructed by including also 
the one-loop contributions which depend on $\hat{\lambda}$, $\hat{\lambda}'$ and the soft parameters $B_i$ 
(see for instance~\cite{Chun:1999bq,Hirsch:2000ef} for a detailed computation of neutrino masses at the one-loop level).
Barring cancellations in $\xi$ the neutrino spectrum turns out to be hierarchical, hence we can impose the relation 
${m}_{{\nu}_{3}} \simeq \sqrt{\Delta m^2_{\rm{atm}}}$, 
thus implying 
\be
\label{eq:mnu3}
\xi \simeq 1.5 \times 10^{-5} {\left(\frac{\sqrt{\Delta m^2_{\rm atm}}}{0.05 \ \mbox{eV}}\right)}^{1/2}
\left(\frac{\tan\beta}{10}\right){\left(\frac{M_{1/2}}{300 \ \mbox{GeV}}\right)}^{1/2} \, .
\ee
Additional constraints on the parameters $\xi_i$ and $\epsilon_i$ can be imposed once other observables, 
such as $\Delta m^2_{\rm sol}$ and the mixing angles, are taken into account.
As it is evident from~\eq{eq:mnu3} the outcome of the analysis of cosmic ray bounds ($\xi < 10^{-6}$) 
implies a suppression of neutrino masses induced by RPV well below the experimental constraints.   
This result is in agreement with the recent analysis of Ref.~\cite{Restrepo:2011rj} 
which rules out bilinear RPV as the mechanism responsible for neutrino masses 
for gravitinos heavier than $1$ GeV. Such low vales of the gravitino mass are not achievable if 
TGM is the dominant mechanism originating sfermion masses (cf.~\eq{eq:gravitinomass}). 
In our model, neutrino masses have to be generated by means of another mechanism 
(see~\app{SO10modelyu} for the discussion on neutrino masses).

\subsubsection*{Gravitino relic density and thermal leptogenesis}

At this point we turn to the relic density of DM. For the range $10^{-(10 \div 11)} \lesssim \xi \lesssim 10^{-(7\div 8)}$ the NLSP decays only into 
SM particles at a much faster rate with respect to the decay into gravitinos. 
The branching ratio between the R-parity conserving decays (into SM particles) 
and the R-parity violating ones (into SM particles and gravitinos),
\be
BR(RPC/RPV) \approx 10^{-8} {\left(\frac{m_{{\chi}_1^0}}{150\ \mbox{GeV}}\right)}^4 {\left(\frac{m_{3/2}}{15\ \mbox{GeV}}\right)}^{-2}{\left(\frac{\tan\beta}{10}\right)}^2
{\left(\frac{\xi}{10^{-10}}\right)}^{-2} \, ,
\ee
indicates that the DM relic density is completely 
determined by its thermal component. 
This can be computed as a function of the reheating temperature $T_{RH}$ in thermal field theory, 
according to the following analytic expression~\cite{Olechowski:2009bd} 
\begin{multline}
\label{eq:thermal-relic}
{\Omega}_{DM}^T{h}^{2}= \\
\left(\frac{m_{3/2}}{10\ \mbox{GeV}}\right)\left(\frac{T_{RH}}{10^9\ \mbox{GeV}}\right)\sum_r y^{\prime}_r\,g^2_r(T_{RH})\left(1+\delta_r\right)
\left(1+\frac{M^2_r(T_{RH})}{2m^2_{3/2}}\right)\ln\left(\frac{k_r}{g_r(T_{RH})}\right) \, ,
\end{multline}
where $r=1,2,3$ and the sum runs over the three components $U(1)_Y$, $SU(2)_L$ and $SU(3)_C$ of the SM gauge group. 
The values of the coefficients $k_r$, $y^{\prime}_r$ and $\delta_r$ can be found for instance in Ref.~\cite{Pradler:2006qh}.
 
\begin{figure}[htbp]
\centering
\includegraphics[width=7 cm, height= 7 cm, angle=360]{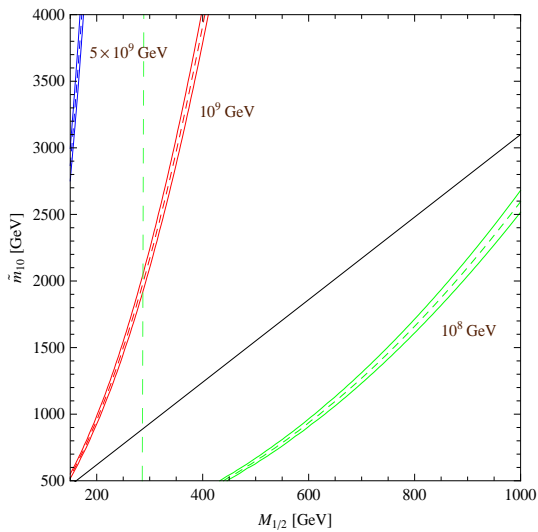}
\caption{\label{OmGrav}Contours of the gravitino relic density in the plane $(M_{1/2},\tilde{m}_{10})$ computed, according to~\eq{eq:thermal-relic}, 
for the values of the reheating temperature reported. 
The green dashed vertical line represents the value of 700 GeV for the gluino mass while the black line represents the 
TGM dominance relation in~\eq{eq:tgmdominance}.}
\end{figure}

As shown in~\fig{OmGrav} the DM relic density matches the cosmological value, depending on the values of 
$M_{1/2}$ and $\tilde{m}_{10}$, for reheating temperatures up to $\sim 10^{9}$ GeV.  
By estimating~\eq{eq:thermal-relic} in the following way
\be
\label{eq:rdestimatefirst}
{\Omega}_{DM}^T{h}^{2} \approx 0.12 \ \left( \frac{T_{RH}}{10^9 \ \rm{GeV}} \right) 
\left( \frac{30 \ \rm{GeV}}{m_{3/2}} \right) \left( \frac{M_{1/2}}{300 \ \rm{GeV}} \right)^{2} \, ,
\ee
and considering the relation in~\eq{eq:gravitinomass}, we can see that the increase in the contribution of TGM to sfermion masses (with respect to standard gauge mediation) coincides with an increase of the reheating temperature needed to fit the cosmological value of the DM relic density. 
Moreover values of the reheating temperature of the order of $10^{9}$ GeV are welcome in the standard  thermal leptogenesis scenarios~\cite{Fukugita:1986hr}. 
We stress that this result is obtained thanks to the natural prediction of TGM of a gravitino mass of the order of 10 GeV. 
On the contrary theories based on standard (loop) gauge mediation predicts much lower reheating temperatures due to lower gravitino masses. 
By rescaling~\eq{eq:rdestimatefirst}, in terms of $\tilde{m}_{10}$ and the gluino mass
\be
\label{eq:rdestimate}
{\Omega}_{DM}^T{h}^{2} \approx 0.12 \ \left( \frac{T_{RH}}{10^9 \ \rm{GeV}} \right) 
\left( \frac{2 \ \rm{TeV}}{\tilde{m}_{10}} \right) \left( \frac{M_3}{700 \ \rm{GeV}} \right)^{2} \, ,
\ee  
we see that a reheating temperature of the order of $10^9$ GeV requires sfermion masses in the multi TeV range.

We conclude by mentioning that, in presence of a leptogenesis mechanism, the theory is subject to additional constraints on the amount of RPV.
Indeed baryon and lepton number violating interactions due to RPV could erase the $B-L$ asymmetry generated by leptogenesis. This can be avoided by requiring that the dangerous processes are not efficient, i.e.~${\Gamma}_{\rm{RPV}} < H$, when the asymmetry is generated. 
The general expression for these rates has been studied in~\cite{Campbell:1990fa,Campbell:1991at}. 
In case of bilinear R-parity violation the baryogenesis bound can be summarized by the condition~\cite{Buchmuller:2007ui} 
\begin{equation}
\epsilon \lesssim 10^{-6} {\left(\frac{\tan\beta}{10}\right)}^{-1} \, ,
\end{equation}
which implies a similar bound on $\xi$, barring cancellations in~\eq{eq:xi}.
\vspace{0.3cm}

The bounds emerging from the cosmological analysis are collectively 
summarized in~\Table{RPVbounds}.
\begin{table}[htdp]
\begin{center}
\begin{tabular}{|c|c|c|}
\hline
Observable & Bound & References \\
\hline 
Proton decay & $\lambda, \lambda^{'}, \lambda^{''} < 10^{-10}$ & \cite{Smirnov:1995ey} \\
BBN (NLSP lifetime) & $\xi > 10^{-(10 \div 11)}$ & \cite{Ishiwata:2008cu} \\
Cosmic rays & $\xi < 10^{-(6 \div 8)}$ & \cite{Choi:2010jt,Vertongen:2011mu} \\
Neutrino masses & $\xi \simeq 10^{-5}$
& \cite{Barbier:2004ez} \\
Baryogenesis & $\epsilon < 10^{-6}$ & \cite{Campbell:1990fa,Campbell:1991at} \\
\hline
\end{tabular}
\end{center}
\caption{Summary of the bounds on the R-parity violating couplings.}
\label{RPVbounds}
\end{table}%

\section{Conclusions}
 
In this paper we have discussed the impact of cosmology on the $SO(10)$ TGM model. 
The new gauge mediation mechanism introduced in~\cite{Nardecchia:2009ew} guarantees the flavor universality of the sfermion masses and peculiar predictions for the supersymmetric spectrum at the LHC. 
The messenger scale is fixed at the GUT scale and this gives also a prediction for the value of the gravitino mass to be of $\mathcal{O}$(10 GeV) 
by assuming the sfermion masses in the TeV range.  Another consequence of the model, under the assumption of TGM dominance (cf.~\eq{eq:tgmdominance}),  is the fact that the NLSP is a Bino-like neutralino.

This scenario is highly disfavored if R-parity is conserved, being the DM overproduced by the NLSP decays. Moreover the gravitino production is accompanied 
by hadronic and/or electromagnetic showers which spoil the predictions of BBN. 

On the contrary the TGM model is naturally feasible in presence of a small amount of R-parity violation. Furthermore the GUT structure of the theory and the constraints from the proton decay motivates the restriction to a scenario of bilinear R-parity violation which allows to describe the relevant phenomenology in terms of a limited number of parameters.
Given the correct amount of RPV,  the NLSP  is allowed to decay much before the onset of BBN without overproducing gravitinos. The latter, at the same time, 
remain stable over cosmological times, being their decay rate doubly suppressed both by the R-parity violating couplings and by the Planck scale. 
Interestingly the small amount of decays which can take place at the present time is in principle detectable by the current cosmic ray experiments such as FERMI-LAT.

The same RPV couplings responsible for the NSLP and gravitino decay could be at the origin of neutrino masses. 
However the value required for these couplings is not compatible with the bounds from cosmic rays, at least for the gravitino masses predicted by TGM.

On the other hand a gravitino with mass of $\mathcal{O}$(10 GeV), combined with a natural TGM spectrum with sfermions in the multi TeV region, allows the DM relic density to match the cosmological value for a reheating temperature of $\mathcal{O}(10^9 \ \rm{GeV})$ relevant for leptogenesis. This improves the situation with respect to the case of the standard (loop) gauge mediation, where the gravitino mass can be at most of $\mathcal{O}$(1 GeV) implying a lower reheating temperature.

\acknowledgments
We thank Ilias Cholis, Maurizio Monaco, Andrea Romanino, Piero Ullio and Robert Ziegler for useful discussions. We also thank Jasper Hasenkamp and Jonathan Roberts for providing us the BBN bounds on the NLSP decays reproduced in~\fig{fig:ben2}.

\appendix

\section{TGM dominance for sfermion masses}
\label{ap:1looprge}

In this appendix we perform an analytical study of the one-loop renormalization group equations (RGEs) for the relevant MSSM parameters in order to investigate the nature of the NLSP and to provide a criterium for the determination of the TGM dominance to sfermion masses. 
Defining $t = \log ( \mu / \mu_0 )$, the running of gaugino masses for $a=1,2,3$ reads
\begin{equation}
\label{eq:1loopgaugino}
M_a(t)=M_{1/2} \frac{g^2_a (t)}{g^2_0} \, ,
\end{equation}
where $g_0$ and $M_{1/2}$ are evaluated at the GUT scale
and $g_a (t)$ obeys the RGE 
\begin{equation}
\label{gaugerun}
g^{-2}_a(t)=g_0^{-2} - \frac{b_a}{8 \pi^2} \, t \, ,
\end{equation} 
with $b_a = (33/5,1,-3)$ being the one-loop MSSM beta-functions.
Assuming flavor universality at the GUT scale
and neglecting the contributions from the Yukawa couplings and the 
A-terms\footnote{We can safely assume that the Yukawa terms give no sizable contribution 
to the sfermions of the first two families. This is not necessarily the case for the third family, 
but since we are interested in a simple analytical estimate of the running effects we stick anyway 
to this approximation.},
the one-loop RGEs for the scalar soft masses can be written as~\cite{Martin:1997ns}
\begin{equation}
\label{1loopsfermion}
16 \pi^2 \frac{d}{dt} \tilde{m}^2_{Q}=-\sum_{a=1,2,3} 8 \, C_a (Q) \, g^2_a M^2_a + \frac{6}{5} Y_Q \, g_1^2 S \, ,
\end{equation}
where $C_a (Q)$ is the quadratic Casimir relative to the representation $Q$ and to the gauge group $a$, and  
the factor $S$ reads 
\begin{equation}
\label{eq:eqS}
S=\tilde{m}^2_{h_u}-\tilde{m}^2_{h_d}  + 3 \left( \tilde{m}^2_{q} - \tilde{m}^2_{\ell} -2  \tilde{m}^2_{u^c} + \tilde{m}^2_{d^c} + \tilde{m}^2_{e^c} \right) \, .
\end{equation} 
Then, by combining~\eq{1loopsfermion} and~\eq{eq:eqS} we obtain 
\begin{equation}
16 \pi^2 \frac{d}{dt} S = \frac{66}{5} g_1^2 S \, , 
\end{equation}
whose integration yields
\begin{equation}
\label{Srun}
S(t) = S_0 \frac{g_1^2 (t)}{g_0^2} \, ,
\end{equation}
where $S_0$ is evaluated at the GUT scale. Given the $SO(10)$ embedding of the matter superfields 
in~\eq{eq:massrelations} one gets 
 $S_0 \equiv S(0)=\tilde{m}^2_{h_u}(0)-\tilde{m}^2_{h_d}(0)$.

We can now integrate~\eq{1loopsfermion} with the help of~\eq{eq:1loopgaugino},~\eq{gaugerun},~\eq{Srun} and the 
GUT boundary conditions in~\eq{eq:2loop}, obtaining
\begin{equation}
\label{eq:1loopsolution}
\tilde{m}^2_Q (t) = A_Q \, \tilde{m}^2_{10} + (B_Q(t) + 2 \, C_Q \, \eta) \, M^2_{1/2} + D_Q (t) \, S_0 \, , 
\end{equation}
where $A_Q$ is equal to 1 for $i=q, u^c, e^c$ and 2 for $i=\ell, d^c$ (cf.~\eq{eq:massrelations}) and $C_Q$ is the 
total SM quadratic Casimir relative to the representation $Q$ (cf.~\eq{casimirs}).
The coefficients $B_Q$ and $D_Q$ parametrize the effects of the running and can be analytically expressed as
\begin{equation}
B_Q(t)= \sum_{a=1,2,3} 2 \, C_a (Q) \frac{1}{b_a} \left(1-\frac{g_a^{4}(t)}{g_0^{4}} \right) 
\quad \text{and} \quad 
D_Q(t) =  \frac{Y_Q}{11} \left(\frac{g_1^{2}(t)}{g_0^{2}} - 1 \right)  \, .
\end{equation}
Evaluating~\eq{eq:1loopgaugino} and~\eq{eq:1loopsolution}  at the TeV scale one obtains
\begin{align}
\label{eq:softms}
&M_1  \simeq 0.4 \, M_{1/2} \, , \\
&M_2  \simeq 0.8 \, M_{1/2} \, , \\
\label{lem3}
&M_3  \simeq 2.5 \, M_{1/2} \, , \\
\label{wcs}
&\tilde{m}^2_q  \simeq  \tilde{m}^2_{10} + (5.2 + 4.2 \, \eta)  \, M^2_{1/2} -0.009  \,  S_0 \, ,  \\ 
\label{mtuc}
&\tilde{m}^2_{u^c}  \simeq   \tilde{m}^2_{10} + (4.8 + 3.2 \, \eta)  \, M^2_{1/2} +0.03  \,  S_0 \, ,  \\ 
\label{mtec}
&\tilde{m}^2_{e^c}  \simeq   \tilde{m}^2_{10} + (0.1 + 3.2 \, \eta)  \, M^2_{1/2} -0.05  \,  S_0 \, ,  \\ 
\label{mtdc}
&\tilde{m}^2_{d^c}  \simeq  2\,  \tilde{m}^2_{10} + (4.7 + 2.8 \, \eta)  \, M^2_{1/2} -0.02  \,  S_0 \, ,  \\ 
\label{mtell}
&\tilde{m}^2_{\ell}  \simeq  2\,  \tilde{m}^2_{10} + (0.5 + 1.2 \, \eta)  \, M^2_{1/2} +0.03  \,  S_0 \, . 
\end{align}

The presence of the contributions due to $M_{1/2}$ makes TGM the leading mechanism for sfermion masses only in some portions of the MSSM parameter space. 
We can define operatively the dominance of TGM by requiring that at least the $50\%$ 
of the low-energy value of the sfermion masses is due to TGM.
Neglecting $S_0$ the worst case scenario in~\eq{wcs} translates into 
\begin{equation}
\tilde{m}_{10}^2 \gtrsim (5.2 + 4.2 \, \eta)  \, M^2_{1/2} 
\, ,
\end{equation}
that is
\begin{equation}
\tilde{m}_{10} \gtrsim  3.1 \, M_{1/2} \, ,
\end{equation}
for $\eta=1$. From this relation it is evident that the NLSP is always the lightest gaugino if TGM is the dominant mechanism generating sfermion masses.

Given $M_{1/2} \simeq 0.4 \,  M_3$ from~\eq{lem3} and taking $M_3 \approx m_{\tilde{g}}$, we arrive to the relation
\begin{equation}
\label{eq:tgmd}
\tilde{m}_{10} \gtrsim 1.2 \, m_{\tilde{g}} \, ,
\end{equation}
that, if substituted into~\eqs{wcs}{mtell}, yields the following bounds on the sfermion masses 
as functions of the gluino mass $m_{\tilde{g}}$:
\begin{align}
\label{mq}
&\tilde{m}_q  >   1.2 \, \textrm {TeV}  \,  \left( \frac{m_{\tilde{g}}}{700 \,  \textrm {GeV} } \right) \, , \\ 
\label{muc}
&\tilde{m}_{u^c}  >  1.2 \, \textrm {TeV}  \,  \left( \frac{m_{\tilde{g}}}{700 \,  \textrm {GeV} } \right) \, , \\ 
\label{mec}
&\tilde{m}_{e^c}  >  1.0 \, \textrm {TeV}  \,  \left( \frac{m_{\tilde{g}}}{700 \,  \textrm {GeV} } \right) \, , \\ 
\label{mdc}
&\tilde{m}_{d^c}  >   1.4 \, \textrm {TeV}  \,  \left( \frac{m_{\tilde{g}}}{700 \,  \textrm {GeV} } \right) \, , \\ 
\label{ml}
&\tilde{m}_{\ell}  >  1.3 \, \textrm {TeV}  \,  \left( \frac{m_{\tilde{g}}}{700 \,  \textrm {GeV} } \right) \, .
\end{align}

\section{Details of the $SO(10)$ model}
\label{SO10model}

In this Appendix we give the details of the $SO(10)$ model presented in~\sect{RPVinTGM}. 
For later convenience let us set the following notation for the SM components of the $SO(10)$ fields relevant for the Yukawa sector
\begin{align}
\label{dec16F}
& 16_F = 
( D^c \oplus L )_{\overline{5}_{-3}} \oplus ( u^c \oplus q \oplus e^c )_{10_{+1}} \oplus (\nu^c)_{1_{+5}} \\ 
\label{dec10F}
& 10_F = 
( D \oplus L^c )_{5_{-2}} \oplus ( d^c \oplus \ell )_{\overline{5}_{+2}} \\
\label{dec16H}
& 16_H =
( T^{16}_d \oplus h^{16}_d )_{\overline{5}_{-3}} \oplus ( \ldots )_{10_{+1}} \oplus ( \ldots )_{1_{+5}} \\
\label{dec16barH}
& \overline{16}_H =
( T^{\overline{16}}_u \oplus h^{\overline{16}}_u )_{5_{+3}} \oplus ( \ldots )_{\overline{10}_{-1}} \oplus ( \ldots )_{1_{-5}} \\
\label{dec10H}
& 10_H =
( T^{10}_u \oplus h^{10}_u )_{5_{-2}} \oplus ( T^{10}_d \oplus h^{10}_d )_{\overline{5}_{+2}}
\end{align}
where a self-explanatory SM notation is employed and the outer subscripts label the $SU(5) \otimes U(1)_X$ origin.
The $SU(2)_L$ doublets decompose as $q = (u \oplus d)$, $\ell = (\nu \oplus e)$, $L = (N \oplus E)$, 
$L^{c} = (E^{c} \oplus N^{c})$, $h_u = (h^{+}_u \oplus h^{0}_u)$ and $h_d = (h^{0}_d \oplus h^{-}_d)$.

\subsection{Symmetry breaking and doublet-triplet splitting}
\label{SO10modelsb}

The set of Higgs fields $54_H \oplus 45_H \oplus 16_H \oplus \overline{16}_H$ is sufficient in order to 
achieve a renormalizable\footnote{With only $45_H \oplus 16_H \oplus \overline{16}_H$ at play the requirement of a supersymmetric vacuum at the GUT scale is such that the little group is $SU(5)$ \cite{Buccella:1981ib,Babu:1994dc,Aulakh:2000sn}. 
In order to reach the SM gauge group one can either relax renormalizability~\cite{Babu:1994dc} or add a $54_H$~\cite{Buccella:1981ib,Aulakh:2000sn}. Since the first option introduces a delicate interplay between 
the GUT and the Planck scale which may be an issue for unification and proton decay (see e.g.~Ref.~\cite{Bertolini:2010yz}), we choose the second option.}
breaking of $SO(10)$ down to the SM (see e.g.~Ref.~\cite{Buccella:2002zt} for the study of the vacuum patterns).
In particular, the SM gauge group is obtained as the intersection of the little groups preserved by the following VEVs:
\begin{align}
& \vev{54_H} \equiv V^{54} \qquad \qquad \ \ SU(4)_C \otimes SU(2)_L \otimes SU(2)_R \, , \\
& \vev{45_H}_{B-L} \equiv V^{45}_{B-L} \qquad\, SU(3)_C \otimes SU(2)_L \otimes SU(2)_R \otimes U(1)_{B-L} \, , \\
& \vev{45_H}_{R} \equiv V^{45}_{R} \qquad \qquad SU(4)_C \otimes SU(2)_L \otimes U(1)_R \, , \\
& \vev{16_H} \equiv V^{16} \qquad \qquad \ \ SU(5) \, .
\end{align}
With the minimal set of Higgs representations at hand we can explicitly check the feasibility of the 
doublet-triplet (DT) splitting. To this end we compute the mass matrices for the doublets ($\mathcal{M}_D$) and the triplets ($\mathcal{M}_T$). 
From $W_H$ in~\eq{WHiggs} we get 
\begin{align}
\label{MD}
\mathcal{M}_D & =
\left(
\begin{array}{cc}
\mu_{10} + \frac{1}{2}\sqrt{\frac{3}{5}}\lambda_{10} V^{54} & \lambda_{16-10} V^{16} \\
\overline{\lambda}_{16-10} V^{16} & \mu_{16} + \lambda_{16} V^{45}_{B-L}
\end{array}
\right) \, , \\
\label{MT}
\mathcal{M}_T & = 
\left(
\begin{array}{cc}
\mu_{10} - \frac{1}{\sqrt{15}} \lambda_{10} V^{54} & \lambda_{16-10} V^{16} \\
\overline{\lambda}_{16-10} V^{16}& \mu_{16} + \lambda_{16} V^{45}_{R}
\end{array}
\right) \, ,
\end{align}
defined, respectively, on the basis $(h^{10}_u, \ h^{\overline{16}}_u) (h^{10}_d, \ h^{16}_d)$ and $(T^{10}_u, \ T^{\overline{16}}_u) (T^{10}_d, \ T^{16}_d)$.
The relevant Clebsch-Gordan coefficients can be found for instance in Ref.~\cite{Fukuyama:2004ps}. 

Two light Higgs doublets, $h_u$ and $h_d$, can be obtained by imposing the minimal fine-tuning condition 
$\det \mathcal{M}_D \sim 0$ in~\eq{MD}, while leaving at the same time the triplets at the GUT scale (cf.~\eq{MT}). 
Working for simplicity in the real approximation the light components read 
\be
h_u = \cos{\theta_u} h^{10}_u + \sin{\theta_u} h^{\overline{16}}_u \, , \qquad
h_d = \cos{\theta_d} h^{10}_d + \sin{\theta_d} h^{16}_d \, ,
\ee
where, taking into account the minimal fine-tuning condition, $\theta_{u,d}$ are fixed in the following way in terms of the superpotential parameters 
\be
\tan\theta_u = - \frac{\lambda_{16-10} V^{16}}{\mu_{16} + \lambda_{16} V^{45}_{B-L}} \, , \qquad 
\tan\theta_d = - \frac{\overline{\lambda}_{16-10} V^{16}}{\mu_{16} + \lambda_{16} V^{45}_{B-L}} \, .
\ee
Notice that in general $\theta_u \neq \theta_d$.
In particular, the projection of 
$v^{10,\overline{16}}_u$ and $v^{10,16}_d$
on the electroweak VEVs 
$v_u \equiv \vev{h_u}$ and $v_d \equiv \vev{h_d}$ 
is
\be
v^{10}_u = v_u \cos{\theta_u} \, , \quad
v^{\overline{16}}_u = v_u \sin{\theta_u} \, , \quad 
v^{10}_d = v_d \cos{\theta_d} \, , \quad
v^{16}_d = v_d \sin{\theta_d} \, .
\ee
Worth of a comment is the fact that the natural (without fine-tuning) implementation of the DT splitting requires the introduction of additional representations.
A solution along these lines, in the context of an $SO(10)$ model of TGM, has been put forward in Ref.~\cite{Nardecchia:2009nh}.

\subsection{Yukawa sector in the pure embedding}
\label{SO10modelyu}

Let us turn now to the Yukawa sector of the model. 
The flavor structure of supersymmetric $SO(10)$ GUTs 
with extended matter sector ($16_F \oplus 10_F$) has been extensively studied in Refs.~\cite{Malinsky:2007qy,Heinze:2010du}. 
On the other hand the mechanism of TGM requires a peculiar embedding of the MSSM fields which must fit into $SU(5)$ representations 
with positive 
$X$-charge
where $SU(5) \otimes U(1)_X \subset SO(10)$.
This is needed in order to guarantee positive sfermion masses\footnote{Strictly speaking 
what one has to require from a phenomenological point of view is that possible negative 
contribution to sfermion masses, originating from a non-pure embedding, are anyway subleading with respect to the 
positive ones~\cite{Nardecchia:2009nh}. For simplicity we stick here to the pure embedding limit.} (cf.~e.g.~\eq{soft1}).

Such an embedding is explicitly shown in~\eqs{dec16F}{dec10F}, with the lower-case fields 
($q$, $u^c$, $d^c$, $\ell$, $e^c$, $\nu^c$)
labeling the MSSM degrees of freedom. 
In order for this to work one has to ensure that the vector-like pairs $D^c \oplus L$ and $D \oplus L^c$ (cf.~again~\eqs{dec16F}{dec10F}) 
pick up a super-heavy mass term, thus decoupling from the low-energy spectrum. 

After the electroweak symmetry breaking the fields with the same unbroken quantum numbers 
mix among themselves. As far as the charged fermions are concerned the superpotential $W_Y$ in~\eq{WYuk} yields 
the following mass matrices 
\be
\label{Mcf}
M_u = Y_{10} v^{10}_u \, , \qquad
M_d =
\left(
\begin{array}{cc}
Y_{10} v^{10}_d & Y_{16} v^{16}_d \\
Y_{16}^T V^{16} & M_{\Delta}
\end{array}
\right) \, , \qquad
M_e =
\left(
\begin{array}{cc}
Y_{10} v^{10}_d & Y_{16} V^{16} \\
Y_{16}^T v^{16}_d & M_{\Lambda}
\end{array}
\right) \, ,
\ee
defined respectively on the basis $(u)(u^c)$, $(d, \ D)(D^c, \ d^c)$ and $(E, \ e)(e^c, \ E^{c})$.
We also defined (see e.g.~Refs.~\cite{Malinsky:2007qy,Heinze:2010du})
\bea
&& M_\Delta \equiv M_{10} + \eta V^{45}_{B-L} - \tfrac{1}{\sqrt{15}} \lambda V^{54} \, , \\
&& M_\Lambda \equiv M_{10} - \eta V^{45}_{R} + \tfrac{1}{2} \sqrt{\tfrac{3}{5}} \lambda V^{54} \, ,
\eea 
where $M_{10}$ and $\lambda$ ($\eta$) are symmetric (antisymmetric) matrices in flavor 
space\footnote{The reason being simply because $10 \otimes 10 = 1_S \oplus 45_A \oplus 54_S$.}. 
Thus, by inspecting~\eq{Mcf}, the decoupling of the vector-like pairs $D^c \oplus L$ and $D \oplus L^c$
is achieved by requiring 
\be
\label{pureembedd}
M_\Delta = M_\Lambda = 0\, .
\ee 
We call this the pure embedding condition which 
gives the desired embedding up to $v_d / V^{16} \ll 1$ corrections.
Given the symmetry properties of the matrices $M_{10}$, $\eta$ and $\lambda$ and the need to 
keep the VEVs of the $45_H$ and $54_H$ switched on for the $SO(10)$ symmetry breaking, 
the pure embedding condition translates into $M_{10} = \eta = \lambda = 0$.

Let us consider now neutrino masses. 
In this case~\eq{WYuk} is responsible for the following Majorana mass matrix
\be
\label{Mnu}
M_\nu =
\left(
\begin{array}{cccc}
0 & Y_{10} v^{10}_u & 0 & Y_{16} V^{16} \\
\cdot & 0 & 0 & Y_{16} v^{16}_d \\
\cdot & \cdot & \lambda w_{+} & M_{\Lambda} \\
\cdot & \cdot & \cdot & \lambda w_{-}
\end{array}
\right) \, ,
\ee
defined on the symmetric basis $(N, \ \nu^c, \nu, \ N^c)$. In~\eq{Mnu} $w_{\pm} \equiv \vev{(1,3,\pm 1)_{54_H}}$ denotes a pair of VEVs 
induced by $W_H$. The contribution to neutrino masses due to the VEV of the scalar triplets goes under the name of type-II seesaw. 

In order to estimate the order of magnitude of the induced VEVs let us consider the following piece of superpotential evaluated on the vacuum
\begin{multline}
\vev{W_H} 
\supset \mu_{54} \left( (V^{54})^2 + w_+ w_- \right) + \lambda_{54} \left( w_+ w_- V^{54} \right) + \lambda_{10} \left( w_+ (v_d^{10})^2 + w_- (v_u^{10})^2 \right) 
\end{multline}
and require the $F$-term conditions $F_{w_{\pm}} = 0$, which gives
\be
w_{\mp} = - \frac{\lambda_{10} \left( v^{10}_{d,u} \right)^2 }{\mu_{54} + \lambda_{54} V^{54}} = \mathcal{O} \left( \frac{M_W^2}{M_G} \right) \, .
\ee
However, as explained above, 
the pure embedding condition requires $\lambda = 0$ so that the type-II seesaw contribution to light neutrino masses vanishes (cf.~\eq{Mnu}). 

Sticking to a pure embedding one can still invoke non-renormalizable operators in order to give a mass to neutrinos through a 
standard type-I seesaw mechanism\footnote{Notice that in the pure embedding one has the $SU(5)$ 
relation $M_d = M_e^T$ (cf.~\eq{Mcf}) which is phenomenologically unacceptable. 
In this respect the presence of non-renormalizable operators is welcome in order to unlock that relation. }. 
Consider for instance the following Planck-suppressed operators
\be
\frac{Y_D}{M_P} 16_F 10_F \overline{16}_H 10_H \supset \frac{Y_D V^{16}}{M_P} \, \ell \, \nu^c h_u^{\overline{16}} \, ,
\ee
and
\be
\frac{Y_N}{M_P} 16_F 16_F \overline{16}_H \overline{16}_H \supset \frac{Y_N (V^{16})^2}{M_P} \, \nu^c \nu^c \, .
\ee
They contribute to the light neutrino mass matrix after integrating out $\nu^c$, yielding
\be
\label{eq:typeI}
m_\nu^I =  
\left(Y_D Y_N^{-1} Y_D^T\right)  (\sin^2\theta_u)^2 \frac{v_u^2}{M_P} \sim \left(Y_D Y_N^{-1} Y_D^T\right) (\sin\theta_u)^2 \ 10^{-5}  \ \text{eV} \, .
\ee
This value is naturally too small and requires a fine-tuning in the Yukawa structure in order to restore the agreement with the experimental data.

One should also keep in mind that in an R-parity breaking scenario there are 
new lepton number violating operators which contribute to neutrino masses as well.   
However the issue of neutrino masses with R-parity violation is tightly correlated with cosmology
and it turns out that the size of the RPV couplings needed by neutrino masses leads, for the range of gravitino masses expected in TGM, to an unacceptable decay rate 
of gravitinos in view of the recent bounds on cosmic rays (cf.~\sect{generalbounds}). 

The bottom line about neutrino masses is that they are naturally too small in the minimal $SO(10)$ model in consideration, 
though it is always possible to fit them with a standard type-I seesaw mechanism (cf.~\eq{eq:typeI}) 
due to the presence of unknown Yukawa structures 
which are not correlated with the charged fermion sector. 

On the other hand it is also easy to understand that by introducing additional representations 
in the game one can fit neutrino masses without too much fine-tuning. An interesting possibility is the introduction of a $54'_H$ 
that couples to $10_H^2$ with strength $\gamma$. If the $54'_H$ does not develop a GUT scale VEV then  
the pure embedding condition in~\eq{pureembedd} is automatically fulfilled with $\gamma \neq 0$, 
yielding a type-II seesaw contribution to neutrino masses~\cite{Frigerio:2008ai,Calibbi:2009wk}.  The latter also provides an interesting leptogenesis mechanism based on the out-of-equilibrium decay of the Higgs triplets. On the other hand it has been pointed out in Ref.~\cite{Frigerio:2008ai} that this mechanism requires an high reheating temperature of at least 
$\mathcal{O} (10^{11} \ \rm{GeV})$, while the cosmological value of the DM relic density is fitted by reheating temperatures pointing towards a standard thermal leptogenesis scenario based on the type-I seesaw (see~\sect{generalbounds} for more details).

\subsection{Origin of the R-parity violating operators}
\label{SO10modelrp}

This last section is devoted to the derivation of the R-parity violating operators in the effective MSSM theory.
Starting from $\delta W_{RPV}$ in~\eq{WRPV} and by projecting the $SO(10)$ representations onto the light 
components (cf.~\eqs{dec16F}{dec10F}) one finds: 
\begin{itemize}
\item A bilinear operator of the type $\mu_i \, \ell_i h_u$, where 
\be
\label{muiGUT}
\mu^i = \cos{\theta_u} \left( \tilde{\mu}_{10}^i - \tilde{\eta}_{10}^i V^{45}_{R} + \tfrac{1}{2} \sqrt{\tfrac{3}{5}} \tilde{\lambda}_{10}^i V^{54}  \right) 
+ \sin{\theta_u} \tilde{\overline{\sigma}}^i V^{16} \, .
\ee
\item Two bilinear operators of the type $\mu_{Ti}^{10} d^{c}_i T^{10}_u$ and $\mu_{Ti}^{\overline{16}} d^{c}_i T^{\overline{16}}_u$, where 
\be
\mu_T^{10i}  = \tilde{\mu}_{10}^i  + \tilde{\eta}_{10}^i V^{45}_{B-L} - \tfrac{1}{\sqrt{15}} \tilde{\lambda}_{10}^i V^{54} 
\qquad \text{and} \qquad
\mu_T^{\overline{16}i} = \tilde{\overline{\sigma}}^i V^{16} \, .
\ee    
\end{itemize}
The triplet bilinears can actually generate effective baryon violating trilinears when combined with the Yukawas 
(see e.g.~\cite{Smirnov:1995ey,Giudice:1997wb}). 
This can be easily seen working at the $SO(10)$ level. Take for instance the terms 
\be
W \supset Y^{ij}_{10} 16^i_F 16^j_F 10_H + Y^{ij}_{16} 16^i_F 10^j_F 16_H + \tilde{\mu}^k 10^k_F 10_H 
+ \tilde{\overline{\sigma}}^k 10^k_F \overline{16}_H \overline{16}_H \, ,  
\ee
where $\tilde{\mu}^k = \left( \tilde{\mu}_{10}^k + \tilde{\eta}_{10}^k \vev{45_H} + \tilde{\lambda}_{10}^k \vev{54_H} \right)$ and by integrating out the pairs $10_H - 10_H$ 
and $16_H - \overline{16}_H$ one gets\footnote{The argument should be formally carried on at the SM level by integrating out the 
heavy triplets.}
\be
\frac{Y^{ij}_{10} \tilde{\mu}^k}{\mu_{10}} 16^i_F 16^j_F 10^k_F + \frac{Y^{ij}_{16} \tilde{\overline{\sigma}}^k}{\mu_{16}} 16^i_F 10^j_F 10^k_F \vev{\overline{16}_H} \, .  
\ee 
After projecting these operators on the MSSM fields, only the second one gives a low-energy contribution, 
leading to the trilinear operator $\left(\lambda^{\prime\prime}_T\right)^{ijk} u^{c}_i d^{c}_j d^{c}_k$ with
\be
\label{lambdaGUT}
\left( \lambda^{\prime\prime}_T \right)^{ijk} = \frac{V^{16}}{\mu_{16}} Y^{ij}_{16} \tilde{\overline{\sigma}}^k 
\, .
\ee
The $SO(10)$ trilinear operator $\tilde{\Lambda}^{ijk} 16_F^i 16_F^j 10_F^k $ has no projection on the light MSSM fields as well
and thus does not contribute to the effective low-energy RPV superpotential.
On the other hand RPV trilinear couplings can arise at the non-renormalizable level from the following operator
\begin{multline}
\label{lambdaNR}
 \frac{\tilde{\Lambda}^{ijk}_{NR}}{M_P} \, 10^i_F 10^j_F 16^k_F \vev{\overline{16}_H} \supset \frac{\tilde{\Lambda}^{ijk}_{NR}}{M_P} 
 \, \overline{5}^i_{10_F} \overline{5}^j_{10_F} 
10^k_{16_F} \vev{1_{\overline{16}_H}} \\
=  \frac{\tilde{\Lambda}^{ijk}_{NR} \, V^{16}}{M_P}  \, \left( \ell_i \ell_j e^{c}_k + 2 \,  d^{c}_i \ell_j q_k + d^{c}_i d^{c}_j u^{c}_k \right) \, .
\end{multline}
Notice that due to the antisymmetry of the $10^k_{16_F}$ in the $SU(5)$ space the interactions in~\eq{lambdaNR} 
are antisymmetric in the first two generation indices: $\Lambda^{ijk}_{NR} = - \Lambda^{jik}_{NR}$.

\bibliography{bibfile}{}
\bibliographystyle{hieeetr}

\end{document}